\documentclass[10pt]{article}

\pdfoutput=1

\usepackage{a4wide}
\usepackage{graphicx}
\usepackage[usenames]{color}
\usepackage{amssymb}

\usepackage{amsmath}
\usepackage{epstopdf}
\usepackage{verbatim}
\usepackage{units}

\usepackage{hyperref}

\usepackage{tabularx}
\newcolumntype{L}[1]{>{\raggedright\arraybackslash}p{#1}}
\newcolumntype{C}[1]{>{\centering\arraybackslash}p{#1}}
\newcolumntype{R}[1]{>{\raggedleft\arraybackslash}p{#1}}

\begin{document}

\begin{center}   
\textbf{\LARGE Basal pressure variations induced by a turbulent flow over a wavy surface}

\vspace*{0.2cm}

P. \textsc{Claudin}$^a$, M. Louge$^b$ and B. \textsc{Andreotti}$^c$
\end{center}

{\small
\noindent
$^a$ {Physique et M\'ecanique des Milieux H\'et\'erog\`enes, UMR 7636 CNRS -- ESPCI Paris -- PSL Research University -- Sorbonne Universit\'e -- Universit\'e de Paris, 75005 Paris, France.}\\
$^b$ {Sibley School of Mechanical and Aerospace Engineering, Cornell University, Ithaca, New York, 14853, USA.}\\
$^c$ {Laboratoire de Physique de l'Ecole Normale Sup\'erieure, UMR 8023 ENS -- CNRS -- Universit\'e de Paris -- PSL Research University, 75005 Paris, France.}
}

\begin{abstract}
Turbulent flows over wavy surfaces give rise to the formation of ripples, dunes and other natural bedforms. To predict how much sediment these flows transport, research has focused mainly on basal shear stress, which peaks upstream of the highest topography, and has largely ignored the corresponding pressure variations. In this article, we reanalyze old literature data, as well as more recent wind tunnel results, to shed a new light on pressure induced by a turbulent flow on a sinusoidal surface. While the Bernoulli effect increases the velocity above crests and reduces it in troughs, pressure exhibits variations that lag behind the topography. We extract the in-phase and in-quadrature components from streamwise pressure profiles and compare them to hydrodynamic predictions calibrated on shear stress data.
\end{abstract}

\vspace*{0.5cm}
\begin{center}
Front. Phys. \textbf{9}, 682564 (2021).\\
\href{https://doi.org/10.3389/fphy.2021.682564}{\texttt{https://doi.org/10.3389/fphy.2021.682564}}
\end{center}
\vspace*{0.5cm}

\section{Introduction}
\label{Intro}

Most natural flows occur on evolving topography. The resulting hydrodynamic variations are described by a linear theory that Jackson and Hunt \cite{JH75} developed for wind profiles over low hills. Their work inspired analyses of laminar \cite{B59,B78,CFKMcLSY82,BHA84,L03} and turbulent \cite{T77a,T77b,S80,BHR81,KM85,TMB87,HLR88,BH98,KSH02,FTh09,FCA10,LC17} flows on shallow bedforms, as recently reviewed by Finnigan et al.~\cite{FAHKOPPRT20}. Flow modulation associated with fluid-structure interactions also drives the dynamics of wind-driven wave generation at a liquid surface \cite{SMcW10,PMR16}, or on compliant thin sheets \cite{HHN80,JAC15,ZWBK17}, leading to the flag instability when a free end is allowed to flap \cite{SZ11}.

Most studies of fluid motion on wavy surfaces have focused on basal shear stress, which drives sediment transport \cite{S36}. As Charru et al~\cite{CAC13} reviewed, coupling the latter to the Jackson and Hunt theory or its variants \cite{KSH02,FCA10} explains the formation of erodible objects like sand ripples and dunes, which owe their initial growth to a basal shear stress peaking upstream of the highest elevation. 

However, basal pressure is also affected by evolving topography. In porous sand beds, streamwise pressure variations produce an internal flow that drives humidity and microscopic particles below the surface \cite{LVMBMTOMDOA10}. With strong enough winds, the resulting pore pressure can also relieve part of the bed weight, thereby facilitating the onset of its erosion~\cite{MTLXB14}. At much larger scale, topography-induced pressure variations are important to atmospheric science, especially mountain meteorology \cite{S79,W00}.

Relatively few experiments conducted in air \cite{M37,K70,WHCWWLC91,GI89,FRBA90,GTD96,CWA13,MTLXB14,LNDCRAFGC21}, water or other liquids \cite{ZCH77,ZH79,AH85,WN92,NMcLW93,NH01,PKAR07,V07,HSPZ17} flowing over wavy surfaces staged harmonic bedforms with low enough ratio of amplitude $\zeta$ and wavelength $\lambda$ to avoid flow detachment. This has made it difficult to compare data with linear theories predicated on small $\zeta/\lambda$. As the experiments of Hanratty, et al. \cite{ZCH77} have shown, a hydrodynamic anomaly occurs when the flow response to topographical variations transitions from laminar to turbulent behavior. This phenomenon is at the origin of an instability that carves rippled or scalloped features on surfaces able to sublimize or dissolve into the fluid \cite{H81,MJ10}, but that disappears when the bed becomes hydrodynamically rough \cite{CDA17}. Recently, such sublimation ripples have been found at the surface of the Martian north polar cap \cite{BCBHMPCPD20}, with a typical wavelength much larger than blue ice ripples found in Antarctica \cite{BRH01}, but still proportional to the viscous length $\nu/u_*$ \cite{T79}, based on kinematic viscosity $\nu$ and shear velocity $u_*$. This anomalous hydrodynamic response is also essential to understand how subaqueous or Martian ripples are superimposed on dunes \cite{DACW19}, acting as a separation of small and large scale bedforms. Therefore, a question is whether streamwise pressure variations are also subject to such anomalous transition.

When fluid flowing on a flat surface reaches an ascending bedform, the narrowing of streamlines raises speed and decreases static pressure, as predicted by energy conservation in the Bernoulli equation. To leading order, this effect is captured by dimensionless coefficients $\mathcal{A}$ and $\mathcal{C}$, which respectively represent the bedform's role on speed and pressure. Because speed rises when pressure decreases, these coefficients have opposite signs, $\mathcal{A}>0$ and $\mathcal{C}<0$. In the ``outer region'' far above the surface, such energy conservation holds. However, in the closer ``inner region'', inertia causes fluid flow to lag changes in the bedform, a phenomenon that is captured by dimensionless coefficients $\mathcal{B}$ and $\mathcal{D}$ that are both positive, and respectively represent effects on shear and normal stress. Overall, fluid inertia causes surface shear stress to lead topographical variations with a positive phase $\arctan(\mathcal{B}/\mathcal{A})>0$, whereas static pressure lags those changes with $\arctan(\mathcal{D}/\mathcal{C})<0$. While others have addressed $\mathcal{A}$ and $\mathcal{B}$~\cite{CAC13,CWA13,LNDCRAFGC21}, this paper focuses on $\mathcal{C}$ and $\mathcal{D}$. We begin with a summary of the theory, which predicts how $\mathcal{C}$ and $\mathcal{D}$ depend on the wavenumber $k = 2\pi/\lambda$ of bed oscillations.

Our objective is to review articles reporting pressure measurements on wavy surfaces subject to a turbulent flow. As we will discuss, existing data \cite{M37,K70,C70,ZCH77,MTLXB14} suggest that the anomalous transition in shear stress may also arise in the pressure response. However, we recognize that the corresponding experiments, which were not designed to address this question, do not support a definitive conclusion. In the context of the anomalous transition, a crucial shortcoming of these experiments is their determination of $u_*$, which may have been approximate. Because the coupling of surface pressure with porous media is relevant to industrial applications and the formation of geophysical bedforms, we hope that this article will inspire future experiments in the dimensionless wavenumber range $10^{-3} \lesssim k\nu/u_* \lesssim 10^{-1}$, where our theory predicts distinct behavior of $\mathcal{D}$ for rough and smooth walls.

\section{Turbulent flow over a wavy bed}
\label{Theory}

Because our main objective is to reanalyze existing data for turbulent flows over wavy beds, this section does not repeat our own derivations of the underlying theory, but rather provides a summary of key quantities and concepts. To account for the hydrodynamic anomaly, the framework of Fourri\`ere, et al.~\cite{FTh09,FCA10} was recently extended, as detailed in \cite{CAC13,CDA17}. We examine a turbulent fluid flow along the $x$ direction, unbounded vertically and driven by a shear stress $\rho u_*^2$ imposed far above the bed. Restricting attention to the linear flow response to bed relief, the elevation $Z(x)$ can be decomposed in Fourier modes. Therefore, without loss of generality, we consider a bed profile of the form
\begin{equation}
Z(x) = - \zeta \cos(kx),
\label{BedCos}
\end{equation}
where $k\zeta$ is a parameter $\ll1$. From Eq.~\ref{BedCos}, troughs reside at $x=0 \mod \left(2\pi\right)$. $z$ is the crosswise distance normal to the reference mean bed elevation. We assume invariance in the spanwise direction $y$ that completes the cartesian coordinate system.

In this framework, hydrodynamics is described by Reynolds-averaged Navier-Stokes equations governing the mean velocity field $u_i$ and pressure $p$. A first order turbulence closure relates the stress tensor $\tau_{ij}$ to the velocity gradient. This closure involves a turbulent kinematic viscosity associated with a mixing length and a mixing frequency representing typical eddy length and time scales \cite{P00}. The mixing frequency is given by the strain rate, and the mixing length $\ell$ depends explicitly on distance from the bed. To account for both the smooth and rough regimes, we adopt a mixing length inspired from van Driest \cite{vD56}
\begin{equation}
\ell=\kappa (z+r \times d-Z) \left\{1-\exp\left[-\frac{(z+s \times d-Z)\sqrt{\tau_{xz}/\rho}}{\nu \mathcal{R}_t}\right]\right\} ,
\label{ellcombo}
\end{equation}
where $\kappa=0.4$ is von K\'arm\'an's constant, $\tau_{xz}$ is the bed shear stress, and $\rho$ is the constant fluid density. $d$ is the sand-equivalent roughness size, from which we define the Reynolds number $\mathcal{R}_d = d u_*/\nu$. The exponential term in Eq.~\eqref{ellcombo} suppresses turbulent mixing within the viscous sub-layer close to the bed. The term $r \times d$ corresponds to the standard Prandtl hydrodynamic roughness $z_0$ extracted by extrapolating the logarithmic law of the wall at vanishing velocity, $u_x = \left(u_*/\kappa\right) \ln \left(z/z_0\right)$. The term $s \times d$ controls the reduction of the viscous layer thickness upon increasing bed roughness. In the rough limit where $d \propto \mathcal{R}_d \to \infty$, the exponential term vanishes and the hydrodynamic anomaly is suppressed altogether. The dimensionless parameters $r \simeq 1/30$ and $s \simeq 1/3$ are calibrated from measurements of velocity profiles over various rough walls \cite{SF09,FS10}.

In Eq.~\eqref{ellcombo}, $\mathcal{R}_t$ is the van Driest transitional Reynolds number. Following Hanratty \cite{H81}, the hydrodynamic anomaly is captured by a spatial relaxation of $\mathcal{R}_t$. In the homogeneous case of a flat bed, it is $\mathcal{R}^0_t \simeq 25$. However, in general, $\mathcal{R}_t$ is not constant but instead trails behind the pressure gradient by a space lag on the order of $\left(\nu/u_*\right)$ that is associated with a thickening of the boundary layer,
\begin{equation}
(a \nu/u_*) \partial_x \mathcal{R}_t = b \mathcal{R}^0_t \nu/(\rho u_*^3) \partial_x (\tau_{xx} -p) - (\mathcal{R}_t - \mathcal{R}^0_t) .
\label{vanDriestEq}
\end{equation}
Charru, et al.~\cite{CAC13} calibrated this additional equation with $a \simeq 2000$ and $b\simeq 35$ by matching theoretical predictions to basal shear stress measurements~\cite{ZCH77,FH88,FRBA90,PKAR07,CWA13}. These predictions were obtained in the regime where the hydrodynamic equations could be linearized with respect to $k\zeta$, and then solved for boundary conditions~\cite{FTh09,FCA10,CDA17}. In this regime, we write the basal shear stress response $\delta \tau_{xz}$ to the bed perturbation~\eqref{BedCos} as
\begin{equation}
\delta \tau_{xz}/\rho u_*^2 = k\zeta \left[-\mathcal{A} \cos(kx) + \mathcal{B} \sin(kx) \right] .
\label{AandB}
\end{equation}
Here, $\rho u_*^2$ is reference shear stress in the flat base state. In Eq.~\eqref{AandB}, the two terms in straight brackets respectively quantify the in-phase and in-quadrature contributions of the response. Both $\mathcal{A}$ and $\mathcal{B}$ have values of order unity \cite{CWA13,LNDCRAFGC21} and are positive, thereby producing a shear stress leading the bed 
elevation. They are weak logarithmic functions of the bed wavenumber, except in the range $10^{-4} \lesssim kz_0 \lesssim 10^{-2}$ where strong variations arise around the hydrodynamic anomaly~\cite{CAC13,CDA17}.

\begin{figure}[p]
\centerline{\includegraphics{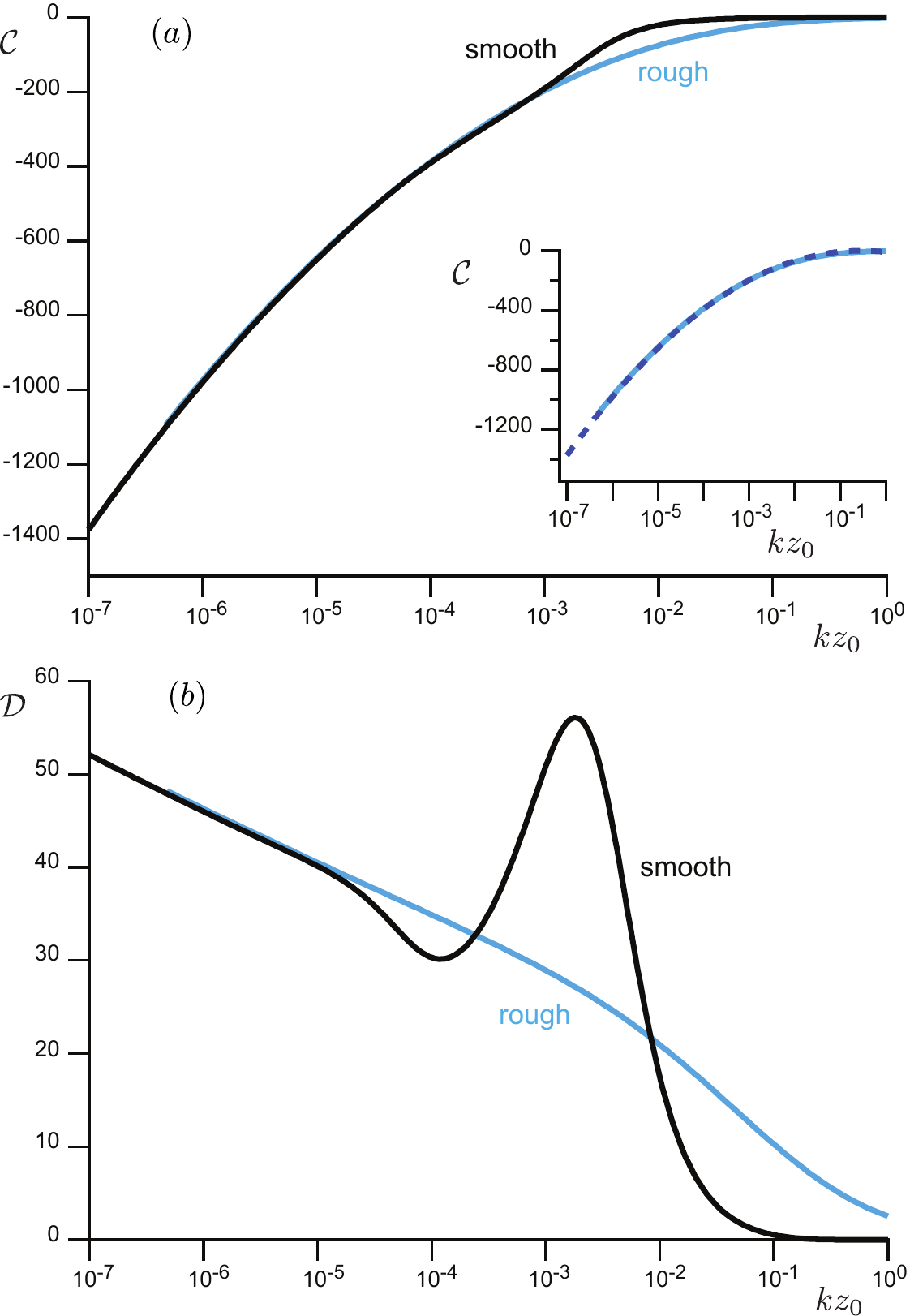}}
\caption{Basal pressure coefficients in terms of the rescaled wavenumber $k \times z_0$ from predictions in~\cite{CDA17}. Panel (a): in-phase component $\mathcal{C}$. Panel (b): in-quadrature component $\mathcal{D}$. Blue lines: rough bed case with roughness $z_0=r \times d$ and $\mathcal{R}_d \gg 1$, calculated from \cite{FTh09,FCA10}. Black lines: smooth bed case $\mathcal{R}_d \to 0$, for which the effective bed roughness in the logarithmic region is proportional to viscous length~\cite{CAC13,CDA17}. Blue and black lines collapse at small wavenumbers adjusting the roughness proportional to the viscous length: $z_0 = \nu /\left(7 u_*\right)$. Inset of panel (a): comparison of the rough case~\cite{FTh09,FCA10} and the quadratic fit $\mathcal{C}=-\left(1/\kappa^2\right) \ln^2 \left(2\pi 
b/kz_0\right)$, with $b \simeq 0.04$ (dashed line).}
\label{CandDtheory}
\end{figure}

Similarly, we write the basal pressure response $\delta p$ in the linear regime as
\begin{equation}
\delta p/\rho u_*^2 = k\zeta \left[-\mathcal{C} \cos(kx) + \mathcal{D} \sin(kx) \right].
\label{CandD}
\end{equation}
Figure~\ref{CandDtheory} shows how $\mathcal{C}$ and $\mathcal{D}$ vary with the dimensionless wavenumber $k \times z_0$ in the rough $\mathcal{R}_d \gg 1$ and smooth $\mathcal{R}_d \to 0$ limits~\cite{FTh09,FCA10,CDA17}. The opposite signs $\mathcal{C} < 0$ and $\mathcal{D} > 0$ mean that, in contrast with shear stress, the phase of pressure modulations are delayed with respect to the bed profile. In addition, because $|\mathcal{C}| \gg \mathcal{D}$, the pressure response is dominated by the Bernoulli effect, in that it is nearly out-of-phase with the topography. Therefore, variations of $\mathcal{C}$ with wavenumber are well captured by approximating the flow as inviscid and irrotational. In this case, the pressure varies as the square of the velocity at a height $\lesssim \lambda$ on the order of the wavelength, which is the only macroscopic scale over which a pressure disturbance is expected to penetrate the flow. From the logarithmic law of the wall, the velocity therefore scales as $\left(u_*/\kappa\right) \ln \left(b \lambda/z_0\right)$, where $b$ is $\lesssim1$. This argument suggests that, in the rough case where complications associated with the hydrodynamic anomaly do not arise, $\mathcal{C}$ should scale as the square of $\ln \left(k z_0 \right)$. As the parabola in Fig.~\ref{CandDtheory}a shows, this approximation indeed conforms well to the theory for the rough case~\cite{FTh09,FCA10}. The smooth case differs from this log-parabolic behavior above $kz_0 \simeq 10^{-3}$ where the anomaly comes into play. While the dependence of $\mathcal{C}$ flattens somewhat at the larger wavenumbers, the anomaly has a more pronounced effect on $\mathcal{D}$, with a distinctive non-monotonic behavior spanning a decade around $kz_0 \simeq 10^{-3}$ (Fig.~\ref{CandDtheory}b).

\section{Pressure measurements over wavy surfaces}
\label{Experiments}

In this section, we compare theoretical predictions to available experimental data. We first outline how to fit the recorded pressure profiles. Then, for each set, we discuss how this procedure yields $\mathcal{C}$, $\mathcal{D}$, and their respective uncertainties.

\subsection{Fitting procedure}
Because the theory is built on a linear analysis of hydrodynamic equations, we restrict attention to data sets with a harmonic pressure response to topographical variations at low $k\zeta \lesssim 0.2$ (Table~\ref{DataTable}). However, as Fig.~\ref{DataZilker} and graphs in Appendix~\ref{AppendixDataAnalysis} indicate, relatively weak non-linearities are apparent. Accordingly, we fit dimensionless pressure response profiles to third-order expansions of the form
\begin{equation}
\delta p/\rho u_*^2 = k\zeta \left[ \Delta_1 \cos(kx - \phi_1) +  \Delta_2 \cos(2 kx - \phi_2) +  \Delta_3 \cos(3 kx - \phi_3) \right],
\label{fit3rdorder}
\end{equation}
but we infer $\mathcal{C}$, $\mathcal{D}$ from the leading order
\begin{eqnarray}
\mathcal{C} & = & - \Delta_1/\sqrt{1 + \tan^2\phi_1},
\label{CfromDeltaandphi} \\
\mathcal{D} & = & \Delta_1 \tan\phi_1/\sqrt{1 + \tan^2\phi_1}.
\label{DfromDeltaandphi}
\end{eqnarray}
For the data sets under consideration, the second and third terms have amplitudes $\Delta_2$ and $\Delta_3 \ll \Delta_1$. As expected, fitting them to first-order alone ($\Delta_1 \ne 0$, $\Delta_2 = \Delta_3 = 0$) does not significantly affect the resulting $\mathcal{C}$ and $\mathcal{D}$. 

Uncertainties in $\mathcal{C}$ and $\mathcal{D}$ depend not only on experimental scatter in $\Delta_1$ and $\phi_1$, but also on how $u_*$ was inferred. Using Eqs.~\eqref{CfromDeltaandphi}-\eqref{DfromDeltaandphi}, we estimate uncertainties due to scatter by carrying out a least-squares regression of data to $\Delta_1$ and $\phi_1$, while assuming that $\Delta_1$ 
and $\phi_1$ are uncorrelated and normally-distributed. Unfortunately, too little information is available to gauge how accurately $u_*$ was established.

\begin{figure}[p]
\centerline{\includegraphics{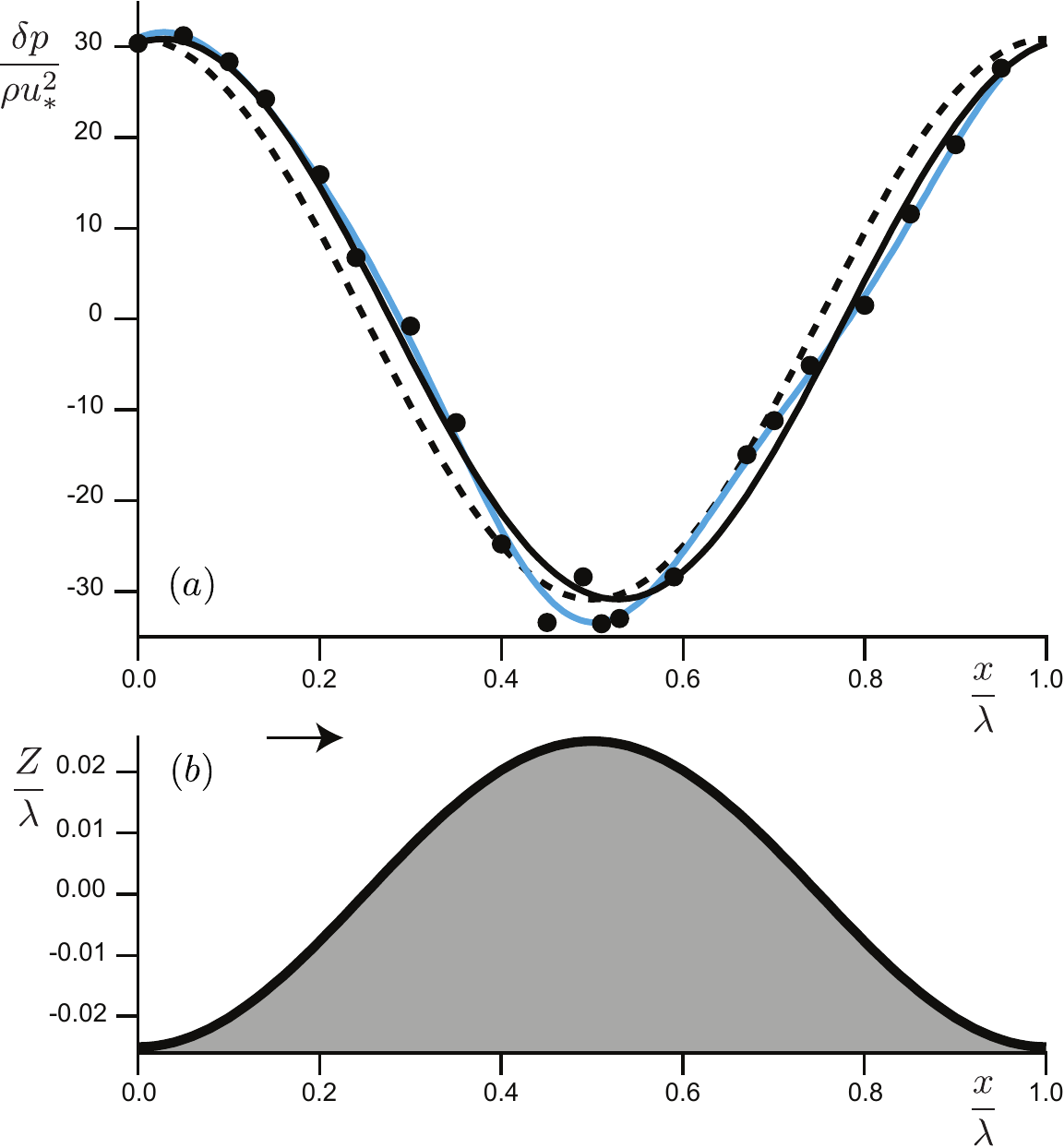}}
\caption{Pressure measurements from \cite{ZCH77}. Table~\ref{DataTable} lists experimental conditions, and ($\mathcal{C}$,$\mathcal{D}$) inferred from ($\Delta_1$,$\phi_1$) using Eqs.~\eqref{CfromDeltaandphi}-\eqref{DfromDeltaandphi}. (a) Basal pressure variations in the flow direction. Filled circles are experimental data. The blue line is a sinusoidal fit to third order (Eq.~\ref{fit3rdorder}). The black line is the harmonic fit $k\zeta \Delta_1 \cos(kx - \phi_1)$. As the dashed line indicates, the in-phase profile $k\zeta \Delta_1 \cos(kx)$ fails to capture data, confirming that pressure variations along the wind lag those of the topography. (b) Bed elevation profile with origin $x=0$ at the trough and flow from 
left to right (arrow).}
\label{DataZilker}
\end{figure}

\subsection{Zilker et al. and Cook}
We first review experiments reported in \cite{ZCH77}, which were used to compare predictions for basal shear stress and related behavior of $\mathcal{A}$ and $\mathcal{B}$ in \cite{CAC13}. Experiments were conducted in a rectangular channel circulating an electrolyte of density $\rho = 1.02 \times 10^3$~kg/m$^3$ and kinematic viscosity $\nu = 8.7 \times 10^{-7}$~m$^2$/s. The bottom of the test section featured interchangeable smooth plexiglas wavy surfaces of fixed $\lambda = 5.08$~cm but different amplitudes $\zeta$. Basal shear stress and static wall pressure were respectively measured with an electrochemical method and small taps. Shear velocity $u_*$ was inferred from measurements of flow rate by integrating the logarithmic law of the wall recorded on a flat surface.

As Fig.~\ref{DataZilker} shows for $k\zeta=0.16$, the pressure profiles exhibit a minimum shortly after the crest. From Eq.~\eqref{fit3rdorder}, we find $\mathcal{C} = -193 \pm 4$ and $\mathcal{D} = 35 \pm 4$. Cook~ \cite{C70} measured additional pressure profiles in a similar setup, albeit using another electrolyte solution with $\rho = 1.05 \times 10^3$~kg/m$^3$ and $\nu = 1.08 \times 10^{-6}$~m$^2$/s. Data at four different flow rates are shown in Fig.~\ref{DataCook}. Pressure amplitude increases with flow velocity, whereas phase shift with respect to bed elevation decreases. $\mathcal{D}$ remains nearly constant $\sim 40$, while $-\mathcal{C}$ rises with $u_*$ (Table~\ref{DataTable}).

\begin{table}[p]
\begin{center}
\renewcommand{\arraystretch}{1.2}
\begin{tabular}{| L{2.5cm} | L{1.5cm} | C{1.5cm} | C{1.5cm} | C{1.5cm} | C{1.5cm} |C{1.5cm} |}
\hline
Experiment	& \multicolumn{6}{l|}{data symbol, bed and fluid properties} \\
\cline{2-7}
			& profiles	& $u_*$ (m/s) 	& $k \nu/u_*$ 	& $\mathcal{C}$ 	& $\mathcal{D}$ 	& $\psi_1$ ($\%$)\\
\hline
\hline
Zilker et al. \cite{ZCH77}	& \multicolumn{6}{l|}{{\Large $\bullet$}, $\lambda = 0.05$~m, $k\zeta = 0.16$, $\nu = 8.7 \times 10^{-7}$~m$^2$/s, $\rho = 1.02 \times 10^3$~kg/m$^3$} \\
\cline{2-7}
			& Fig. 2a	& $0.028$		& $0.0038$	& $ -193 \pm 4$	& $35 \pm 4$	& $2.8 \pm 0.4$ \\
\hline
\hline
Cook \cite{C70}		& \multicolumn{6}{l|}{{\Large $\circ$}, $\lambda = 0.05$~m, $k\zeta = 0.16$, $\nu = 1.08 \times 10^{-6}$~m$^2$/s, $\rho = 
1.05 \times 10^3$~kg/m$^3$} \\
\cline{2-7}
			& Fig. 4a	& $0.030$		& $0.0045$	& $ -140 \pm 3$	& $41 \pm 3$	& $4.5 \pm 0.4$ \\
			& Fig. 4b	& $0.062$		& $0.0022$	& $ -217 \pm 3$	& $42 \pm 3$	& $3.0 \pm 0.3$ \\
			& Fig. 4c	& $0.076$		& $0.0018$	& $ -231 \pm 2$	& $37 \pm 2$	& $2.5 \pm 0.2$ \\
			& Fig. 4d	& $0.079$		& $0.0017$	& $ -246 \pm 2$	& $41 \pm 2$	& $2.6 \pm 0.1$ \\
\hline
\hline
Motzfeld \cite{M37}		& \multicolumn{6}{l|}{{\large $\blacktriangle$}, $\lambda = 0.3$~m, $k\zeta = 0.16$, $\nu = 1.5 \times 10^{-5}$~m$^2$/s, $\rho = 1.2$~kg/m$^3$} \\
\cline{2-7}
			& Fig. 5	& $0.69$		& $0.0005$	& $-377 \pm 22$	& $13 \pm 24$	& $0.5 \pm 
1.0$ \\
\hline
\hline
Kendall \cite{K70}		& \multicolumn{6}{l|}{$\blacksquare$, $\lambda = 0.1$~m, $k\zeta = 0.2$, $\nu = 1.5 \times 10^{-5}$~m$^2$/s, $\rho = 1.2$~kg/m$^3$} \\
\cline{2-7}
			& Fig. 6a	& $0.13$		& $0.0072$	& $-135 \pm 1$		& $51 \pm 1$	& $5.7 \pm 
0.2$ \\
			& Fig. 6b	& $0.21$		& $0.0044$	& $-166 \pm 1$		& $46 \pm 1$	& $4.3 \pm 
0.1$ \\
			& Fig. 6c	& $0.30$		& $0.0031$	& $-193 \pm 1$		& $41 \pm 1$	& $3.3 \pm 
0.1$ \\
			& Fig. 6d	& $0.39$		& $0.0024$	& $-225 \pm 1$		& $38 \pm 1$	& $2.6 \pm 
0.1$ \\
\hline
\hline
Musa et al. \cite{MTLXB14}	& \multicolumn{6}{l|}{{\Large $\diamond$}, $\lambda = 0.1$~m, $k\zeta = 0.19$, $\nu = 1.5 \times 10^{-5}$~m$^2$/s, $\rho = 1.2$~kg/m$^3$} \\
\cline{2-7}
			& Fig. 7a	& $0.16$		& $0.0058$	& $-225 \pm 11$	& $78 \pm 13$	& $5.3 \pm 1.1$ \\
			& Fig. 7b	& $0.33$		& $0.0028$	& $-284 \pm 11$	& $59 \pm 13$	& $3.2 \pm 0.8$ \\
			& Fig. 7c	& $0.55$		& $0.0017$	& $-256 \pm 10$	& $38 \pm 12$	& $2.3 \pm 0.8$ \\
			& Fig. 7d	& $0.76$		& $0.0012$	& $-241 \pm 10$	& $45 \pm 12$	& $2.9 \pm 0.9$ \\
			& Fig. 7e	& $0.95$		& $0.0010$	& $-258 \pm 11$	& $48 \pm 13$	& $2.9 \pm 0.9$ \\
			& Fig. 7f	& $1.21$		& $0.0008$	& $-252 \pm 11$	& $50 \pm 12$	& $3.1 \pm 0.9$ \\
\hline
\end{tabular}
\end{center}
\caption{Experimental conditions and values of $\mathcal{C}$ and $\mathcal{D}$ inferred from harmonic fits of pressure profiles using Eqs.~\eqref{CfromDeltaandphi}-\eqref{DfromDeltaandphi}. Last column: downwind shift relative to bed wavelength $\psi_1 = \phi_1/(2\pi)$, calculated from $\tan\left( 2\pi \psi_1 \right) = -\mathcal{D}/\mathcal{C}$.}
\label{DataTable}
\end{table}

\subsection{Motzfeld and Kendall}
We analyzed experiments performed in wind tunnels over smooth solid sinusoidal waves. The oldest work is Motzfeld's \cite{M37}, who staged four different bed profiles carved in plaster and varnished. To stay within reach of the linear assumption, we only exploited his data for the smallest amplitude (his `model I' with $k\zeta = 0.16$ and $\lambda = 0.3$~m). The corresponding pressure profile and fits are shown in Fig.~\ref{DataMotzfeld}. Equation~\eqref{fit3rdorder} yields a relatively precise $\mathcal{C} = -377 \pm 22$, but a less accurate $\mathcal{D} = 13 \pm 24$, which hints at pressure variations almost in phase with the bed. 

The other wind tunnel data are Kendall's \cite{K70}, who studied turbulent flows over mobile and immobile waves on a rubber surface. We only considered his immobile sinusoidal bed ($k\zeta = 0.2$, $\lambda = 0.1$~m) at different wind velocities. Profiles are shown in Fig.~\ref{DataKendall}. Consistent with the results of \cite{C70}, Kendall's~\cite{K70} pressure amplitude increases with flow velocity, while phase with respect to bed elevation decreases.  

Motzfeld~\cite{M37} and Kendall~\cite{K70} both inferred shear velocity from logarithmic fits of vertical velocity profiles. However, Kendall fitted $z_0$ and $u_*$ separately. Because hydrodynamic roughness on a smooth 
wall is correlated with shear velocity, we recalculated Kendall's $u_*$ using $z_0 \simeq \nu/\left(7 u_*\right)$.

\subsection{Musa et al.}
\begin{figure}[p]
\centerline{\includegraphics{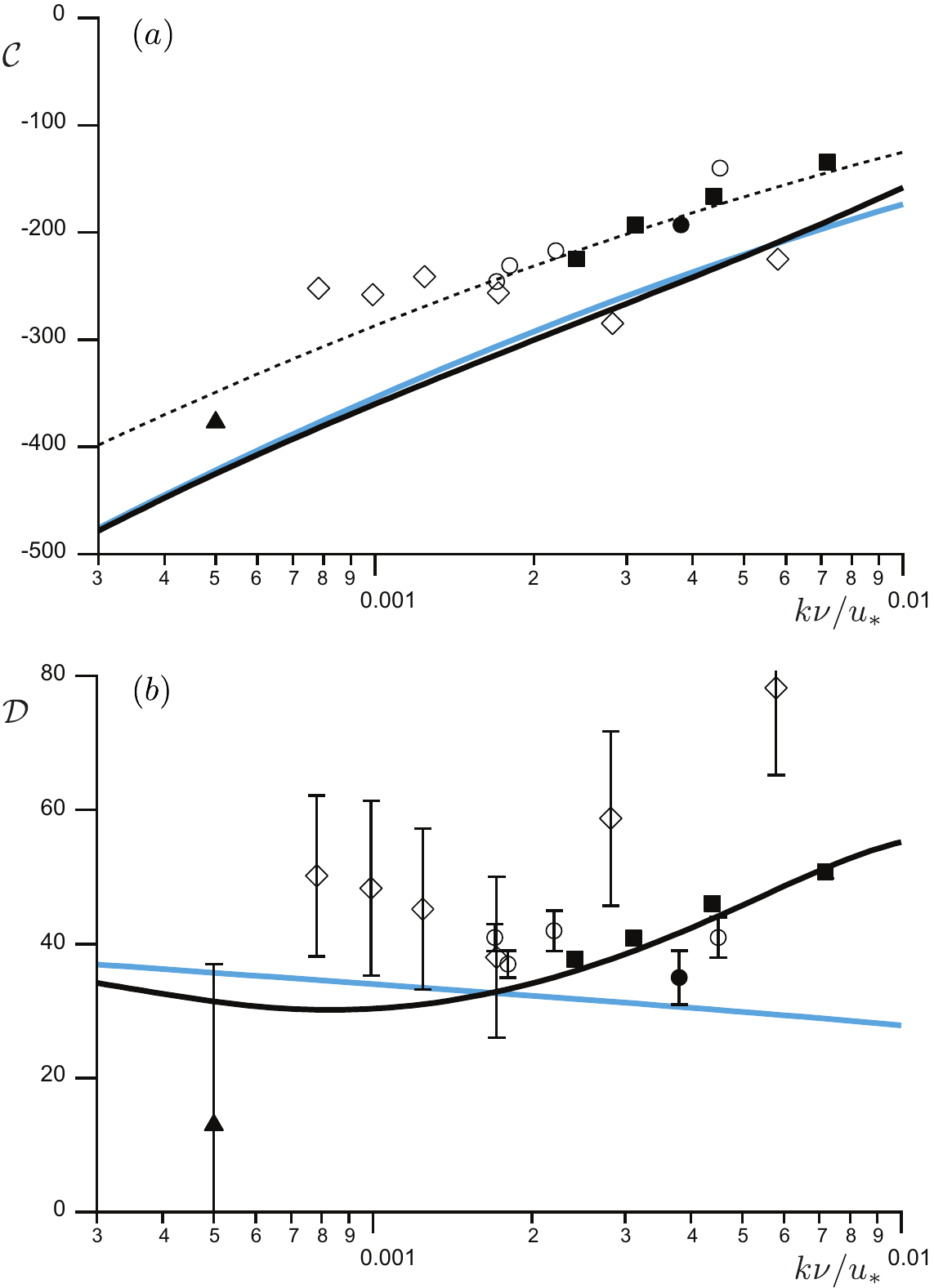}}
\caption{Pressure coefficients $\mathcal{C}$ (a) and $\mathcal{D}$ (b) vs 
$k\nu/u_*$. Black lines are theoretical predictions for a smooth bed (Fig.~\ref{CandDtheory}). The dotted line in (a) is a fit of the form $\mathcal{C}=-\left(1/\kappa^2\right) \ln^2 \left(2\pi b/kz_0\right)$ with $b = 0.02$. Blue lines: theoretical predictions for a rough bed, with $k\nu/u_* = 7 k z_0$.  Symbols, see Table~\ref{DataTable}.}
\label{CandDwithdata}
\end{figure}

Musa et al. \cite{MTLXB14} also acquired data on sinusoidal, smooth, rigid walls in the wind tunnel ($\lambda = 0.1$~m), but their objective was to record pore pressure within a permeable rigid material mimicking a sand bed. We only consider their ripple of smallest amplitude ($k\zeta = 0.19$), which they deployed at six different wind speeds. From a solution of the Laplace equation governing pore pressure, Musa et al. fitted $\delta p$ at the surface to their pore pressure measurements at $45$ locations within the ripple. (In the nomenclature of their Eq. (9), $p_1 \cos(2\pi x/\lambda - \phi_1)$ is equivalent to our $\delta p$). They then inferred shear velocities by fitting vertical wind profiles to the logarithmic law of the wall using an aerodynamic roughness proportional to the viscous length. For consistency with other data reviewed here, we adopt $z_0 = \nu/\left(7u_*\right)$, which differs slightly from their fit of $z_0$ in the smooth limit. Surface pressure profiles and fits to Eq.~\eqref{fit3rdorder} are shown in Fig.~\ref{DataMusa} for their six values of $u_*$. As the open diamonds in Fig.~\ref{CandDwithdata} suggest, $-\mathcal{C}$ rose slowly with $u_*$ and $\mathcal{D}$ was non-monotonic.

\section{Discussion and concluding remarks}
\label{conclusion}

Table~\ref{DataTable} summarizes conditions of all available experiments on smooth walls with nearly harmonic response, and the resulting $\mathcal{C}$ and $\mathcal{D}$. First- and third-order fits of experimental pressure profiles yield similar values for these quantities, thereby demonstrating robustness of the fitting procedure. Uncertainties are relatively small, except for $\mathcal{D}$ from Motzfeld and from Musa, et al.

Figure~\ref{CandDwithdata} shows corresponding variations with the rescaled wavenumber $k\nu/u_*$ and, consistently with experimental conditions, compares them with theoretical predictions in the smooth case~\cite{FTh09,FCA10,CDA17}, which rely on a calibration of the hydrodynamic equations and the relaxation framework of Hanratty~\cite{H81} on the streamwise evolution of basal shear stress~\cite{CAC13}. While these preditions capture the correct trend, they clearly underpredict $\mathcal{C}$. A Bernoulli-like approximation (dotted line in Fig.~\ref{CandDwithdata}a) is more faithful to the data, but it requires $b \sim 0.02$, which is lower than the value used in the inset of Fig.~\ref{CandDtheory} by a factor of $2$. We attribute the discrepancy to challenges in extracting $u_*$ from experiments.

At first glance, Fig.~\ref{CandDwithdata} suggests that $\mathcal{D}$ is nearly constant within experimental error. However, its trend vs $k\nu/u_*$ hints at the presence of a local minimum from the hydrodynamic anomaly for a smooth wall (black curve in Fig.~\ref{CandDtheory}). Nonetheless, the precision in $\mathcal{D}$ is not yet sufficient to distinguish this behavior from that of a rough wall (blue curve).

Overall, $\mathcal{C}$ and $\mathcal{D}$ exhibit dispersion among different experiments. One reason is inconsistent ways to evaluate $u_*$, which affects both axes in Fig.~\ref{CandDwithdata} by rescaling stresses with $\rho u_*^2$ and lengths with $\nu/u_*$. Because velocity fluctuations or Reynolds stresses close to the bed are difficult to measure, $u_*$ was inferred from velocity profiles or flow rates using the law of the wall. In addition, it is open to question how flows driven by a pressure gradient, such as those in wind tunnels or pipes, can be quantitatively compared to a theoretical framework where shear stress is imposed. In this context, the Bernoulli-like approximation suggests that the velocity very close to the bed at an altitude $\simeq b\lambda$ is a better proxy for an effective $u_*$ than the average flow velocity.

A second issue is non-linear effects. Weakly non-linear developments \cite{AFOKMC09,FTh09} and measurements \cite{FCA10} suggest that $k\zeta = 0.2$ is an upper bound for validity of the linear theory. In recorded pressure profiles, we clearly discern weakly non-linear effects, especially in Musa et al. (Fig.~\ref{DataMusa}), whose pressure is lower than expected on bed crests and troughs, although this effect may be due to an interaction with the porous bed underneath~\cite{MTLXB14}. In addition, non-linearities also raise the effective bed roughness on a scale comparable to $\lambda$, with a first corrective term in $(k\zeta)^2$ \cite{FTh09}. This further complicates an estimation of the relevant experimental shear velocity.

These observations call for more measurements, particularly in the range $10^{-3} \lesssim k\nu/u_* \lesssim 10^{-1}$ that resolves the peak of the hydrodynamic anomaly. For air at ordinary wind speed, for example $u_* \simeq 0.5$~m/s, this implies a wavy bed with $2$~mm$< \lambda < 20$~cm. For more gentle winds, the smaller wavelength could rise to $1$~cm. Larger $\lambda$ could be staged with oils of larger viscosity. Data at significantly smaller wavenumbers would require a natural wavy surface such as a sand dune. An example is the hump studied in \cite{CWA13}, where $\lambda \simeq 40$~m and $kz_0 \simeq 10^{-5}$ in the rough limit. Here, the theory predicts $\mathcal{C} \simeq -700$ and $\mathcal{D} \simeq 40$, i.e. a phase shift $\phi_1 \simeq 0.06$ rad, corresponding to a distance $\phi_1/k \simeq 0.4$~m downwind of the crest. If small pressure differences could be reliably recorded over relatively long distances, such spatial phase lag could also be measured.

Finally, DNS or LES simulations would also constitute another source of data, since runs could be performed with strictly imposed values of  $u_*$~\cite{dALB97,HS99}, thereby mirroring the theoretical approach. Unfortunately, simulations of Maa\ss\ and Schumann \cite{MS94,MS96} or those of Salvetti et al. \cite{SDB01} involve amplitudes too large to avoid non-linear effects arising at $k\zeta \gtrsim 0.1$. In both experimental and numerical investigations, $k\nu/u_*$ is typically adjusted at fixed wavenumber using different winds. However, investigating the role of $k$ under constant flow is equally valuable, perhaps again with DNS, to gain a deeper understanding of the hydrodynamic anomaly. At present, the relaxation closure inspired from Hanratty~\cite{H81} is convenient. However, the interplay between a wavy bed and modulation of the viscous sublayer remains an open problem.

The evolution of pressure on geophysical bedforms such as sand ripples creates an internal seepage flow that brings nutrients to the liveforms they shelter~\cite{MGCJH07}, and it provides a mechanism for the accumulation of moisture or dust within them~\cite{LVMBMTOMDOA10}. The phase lag that is proportional to $\arctan(\mathcal{D}/\mathcal{C}) < 0$ also induces surface variations that future research could relate to complex drying patterns that are observed in sand seas~\cite{LVOXHR13}. More generally, pressure variations affect phase change and thermodynamics. Pressure is also a stress scale that may alter the rheology of dense granular flows and suspensions in subtle ways, for example by altering the opening and closing of contacts among grains~\cite{TZGNBC04}, especially when the continuous phase is a liquid~\cite{AFP13}. Although initially motivated by wind flows over ripple-like bed oscillations, the pressure effects that we have reviewed could then be relevant to subjects as diverse as dissolution and sublimation patterns \cite{MJ10,CDA17,CJVPA15}, granular soil liquefaction \cite{AFP13}, droplets and aerosols production \cite{Vi07} and cloud formation over larger scale topography \cite{S79,W00}.

\vspace*{0.3cm}

\noindent
\rule[0.1cm]{3cm}{1pt}

We thank F. Charru, O. Dur\'an and A. Fourri\`ere for fruitful discussions. This research was supported in part by the National Science Foundation 
under Grant No. NSF PHY-1748958, since it was initiated at KITP (Santa Barbara, USA) during the conference on Particle-Laden Flows in Nature (16-22 Dec. 2013), as part of the program `Fluid-mediated particle transport in geophysical flows'. This work has benefited from the financial support of the Agence Nationale de la Recherche, grant `Zephyr' No. ERCS07\underline{\ }18. MYL's contribution was made possible by the support of NPRP grant 6-059-2-023 from the Qatar National Research Fund.


\newpage

\appendix
\section{Data analysis}
\label{AppendixDataAnalysis}

This appendix shows our fitting procedure for all data sets. Figures adopt the same symbols and lines as Fig.~\ref{DataZilker}a. Panels show different shear velocities, as specified in Table~\ref{DataTable}.

\begin{figure}[h]
\centerline{\includegraphics[width=0.9\linewidth]{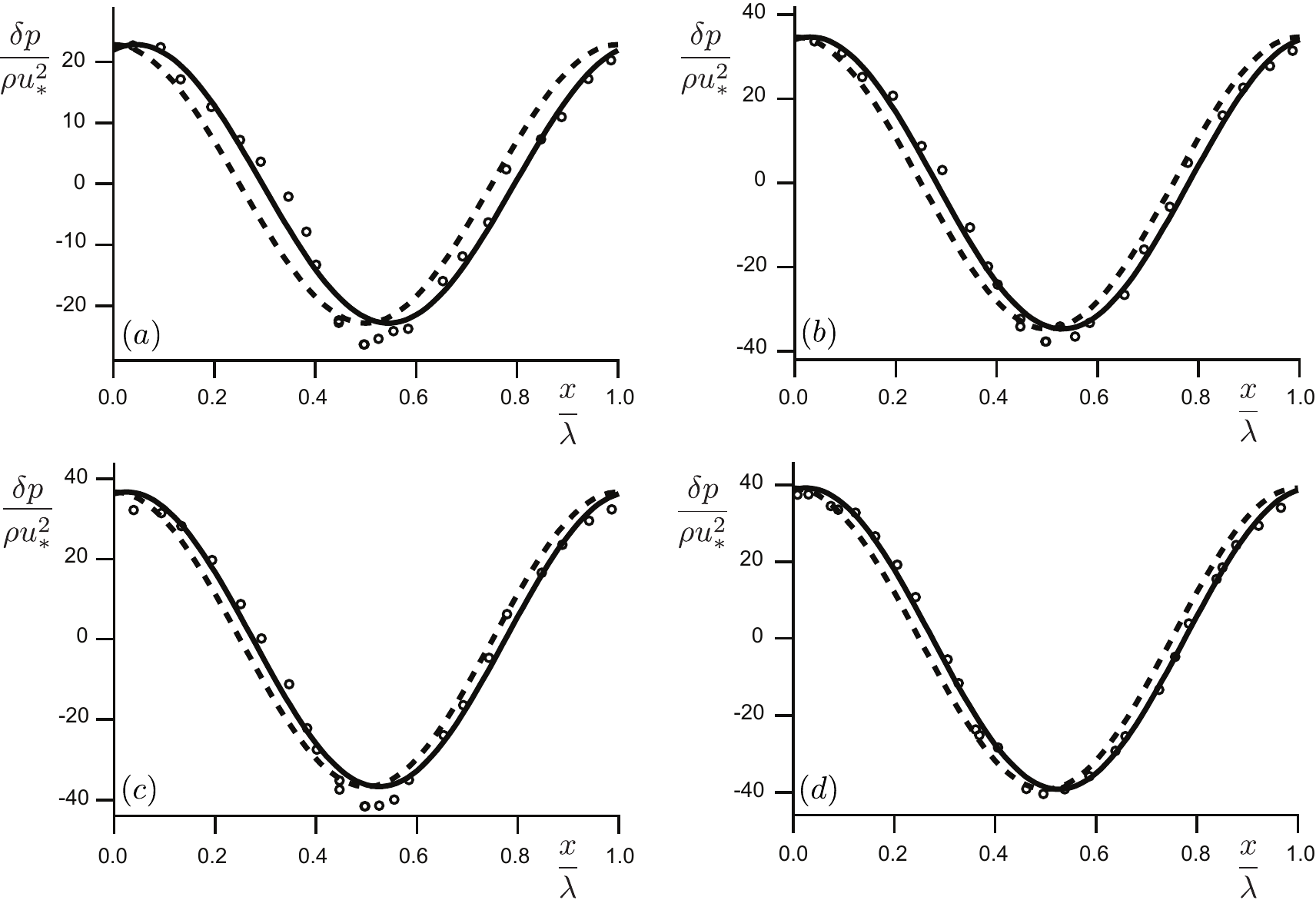}}
\caption{Symbols: pressure measurements from Cook et al. \cite{C70}. Lines, see Fig.~\ref{DataZilker}a.}
\label{DataCook}
\end{figure}

\begin{figure}[t]
\centerline{\includegraphics[width=0.47\linewidth]{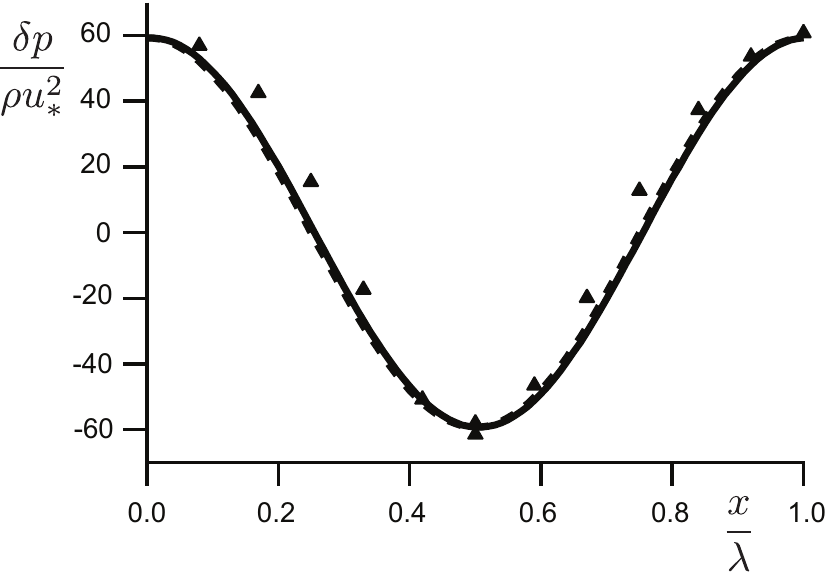}}
\caption{Symbols: pressure measurements from Motzfeld \cite{M37}. Lines, see Fig.~\ref{DataZilker}a.}
\label{DataMotzfeld}
\end{figure}

\begin{figure}[b]
\centerline{\includegraphics[width=0.9\linewidth]{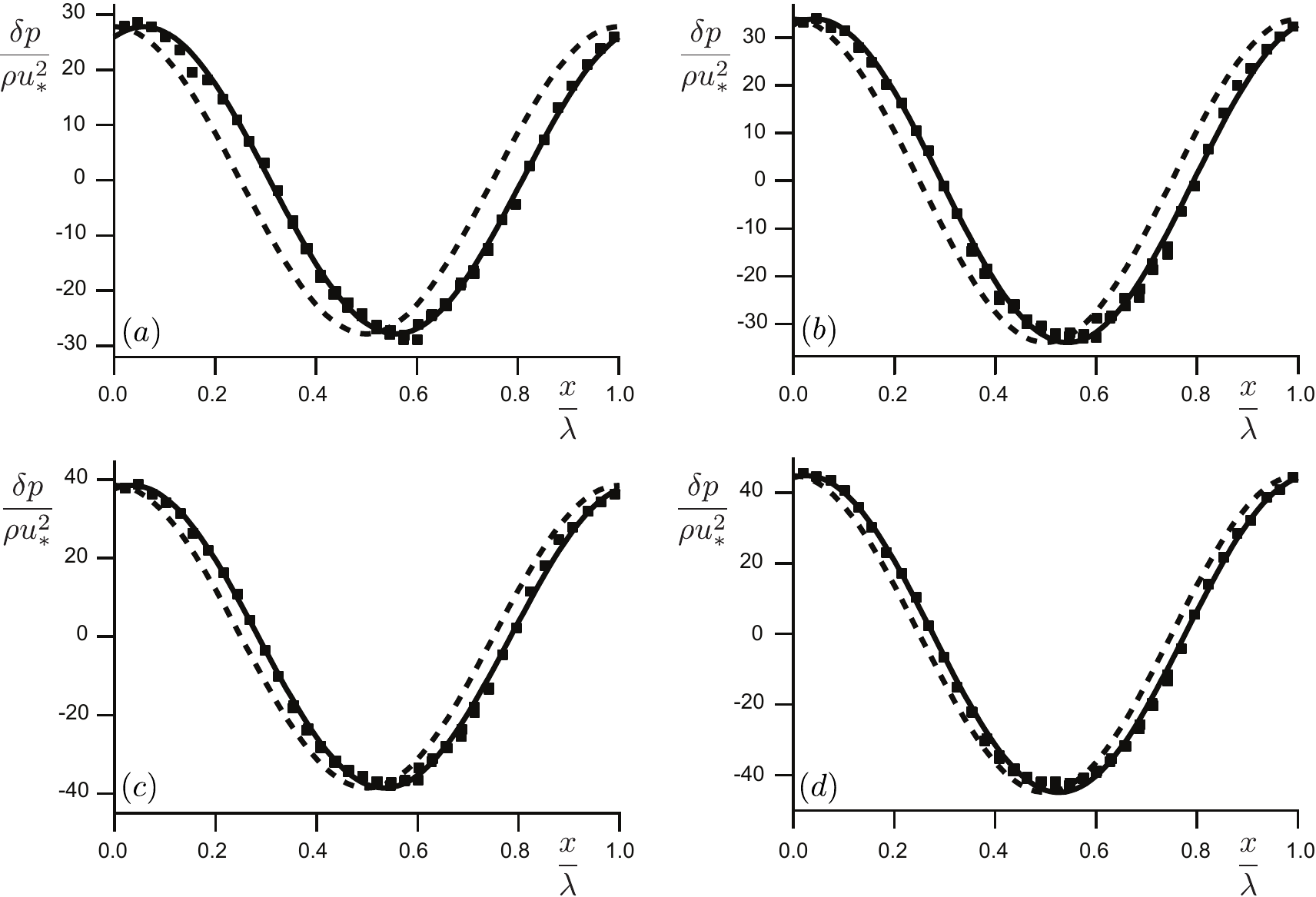}}
\caption{Symbols: pressure measurements from Kendall \cite{K70}. Lines, see Fig.~\ref{DataZilker}a.}
\label{DataKendall}
\end{figure}

\begin{figure}[t]
\centerline{\includegraphics[width=0.9\linewidth]{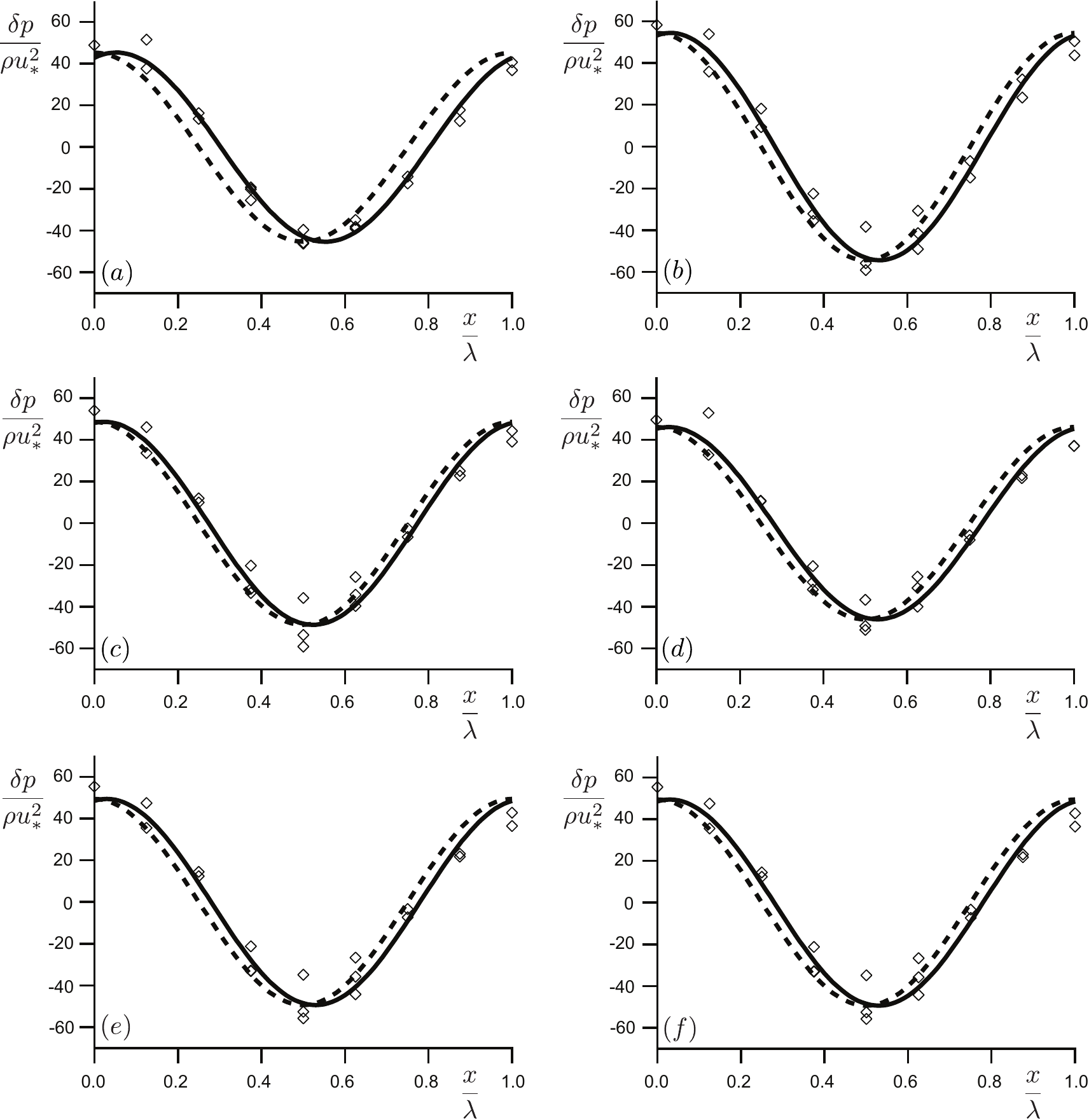}}
\caption{Symbols: surface pressure values extrapolated from pore pressure 
measurements within the porous medium by Musa et al. \cite{MTLXB14}. Lines, see Fig.~\ref{DataZilker}a.}
\label{DataMusa}
\end{figure}


\begin{thebibliography}{}

\bibitem{JH75}
P.S. Jackson and J.C.R. Hunt,
Turbulent wind flow over a low hill.
Q. J. R. Meteorol. Soc. \textbf{101}, 929-955 (1975).

\bibitem{B59}
T.B. Benjamin,
Shearing flow over a wavy boundary.
J. Fluid Mech. \textbf{6}, 161-205 (1959).

\bibitem{B78}
G.L. Bordner,
Nonlinear analysis of laminar boundary layer flow over a periodic wavy surface.
Phys. Fluids \textbf{21}, 1471-1464 (1978).

\bibitem{BHA84}
J. Buckles, T.J. Hanratty and R.J. Adrian,
Turbulent flow over large-amplitude wavy surfaces.
J. Fluid Mech. \textbf{140}, 27-44 (1984).

\bibitem{CFKMcLSY82}
E.A. Caponi, B. Fornberg, D.D. Khight, J.W. McLean, P.G. Saffman and H.C. 
Yuen,
Calculations of laminar viscous flow over a moving wavy surface.
J. Fluid Mech. \textbf{124}, 247-262 (1982).

\bibitem{L03}
P.-Y. Lagr\'ee,
A triple deck model of ripple formation and evolution.
Phys. Fluids \textbf{15}, 2355-2368 (2003).

\bibitem{BH98}
S.E. Belcher and J.C.R. Hunt,
Turbulent flow over hills and waves.
Annu. Rev. Fluid Mech. \textbf{30}, 507-538 (1998).

\bibitem{BHR81}
R.E. Britter, J.C.R. Hunt and K.J. Richards,
Air flow over a two-dimensional hill: studies of  velocity speed-up, roughness effects and turbulence.
Q. J. R. Meteorol. Soc. \textbf{107}, 91-110 (1981).

\bibitem{FTh09}
A. Fourri\`ere,
Morphodynamique des rivi\`eres, s\'election de largeur, rides et dunes (River morphodynamics : Width selection, ripples and dunes). Ph.D. thesis, Universit\'e Paris Diderot (2009).\\
\href{https://pastel.archives-ouvertes.fr/pastel-00005562}{\texttt{https://pastel.archives-ouvertes.fr/pastel-00005562}}

\bibitem{FCA10}
A. Fourri\`ere, P. Claudin and B. Andreotti,
Bedforms in a turbulent stream: formation of ripples by primary linear instability and of dunes by non-linear pattern coarsening.
J. Fluid Mech. \textbf{649}, 287-328 (2010). 

\bibitem{HLR88}
J.C.R. Hunt, S. Leibovich and K.J. Richards,
Turbulent shear flows over low hills.
Q. J. R. Meteorol. Soc. \textbf{114}, 1435-1470 (1988).

\bibitem{KM85}
N. Kobayashi and O.S. Madsen,
Turbulent flows over a wavy bed.
J. Geophys. Res. \textbf{90}, 7323-7331 (1985).

\bibitem{KSH02}
K. Kroy, G. Sauermann and H.J. Herrmann,
Minimal model for aeolian sand dunes.
Phys. Rev. E \textbf{66}, 031302 (2002).

\bibitem{LC17}
P. Luchini and F. Charru,
Quasilaminar regime in the linear response of a turbulent flow to wall waviness.
Phys. Rev. Fluids \textbf{2}, 012601(R) (2017).

\bibitem{S80}
R.I. Sykes,
An asymptotic theory of incompressible turbulent boundary-layer flow over 
a small bump.
J. Fluid Mech. \textbf{101}, 647-670 (1980).

\bibitem{T77a}
P.A. Taylor,
Some numerical studies of surface boundary-layer flow above gentle topography.
Boundary-Layer Meteorol. \textbf{11}, 439-465 (1977).

\bibitem{T77b}
P.A. Taylor,
Numerical studies of neutrally stratified planetary boundary-layer flow above gentle topography.
Boundary-Layer Meteorol. \textbf{12}, 37-60 (1977).

\bibitem{TMB87}
P.A. Taylor, P.J. Mason and E.F. Bradley,
Boundary-layer flow over low hills.
Boundary-Layer Meteorol. \textbf{39}, 107-132 (1987).

\bibitem{FAHKOPPRT20}
J. Finnigan, K. Ayotte, I. Harman, G. Katul, H. Oldroyd, E. Patton, D. Poggi, A. Ross and P. Taylor,
Boundary-layer flow over complex topography.
Boundary-Layer Meteorol. \textbf{177}, 247-313 (2020).

\bibitem{PMR16}
A. Paquier, F. Moisy and M. Rabaud,
Viscosity effects in wind wave generation.
Phys. Rev. Fluids \textbf{1}, 083901 (2016).

\bibitem{SMcW10}
P.P. Sullivan and J.C. McWilliams,
Dynamics of winds and currents coupled to surface waves.
Annu. Rev. Fluid Mech. \textbf{42}, 19-42 (2010).

\bibitem{HHN80}
R.J. Hansen, D.L. Hunston, and C.C. Ni,
An experimental study of flow-generated waves on a flexible surface.
J. Sound Vib. \textbf{68}, 317-334 (1980).

\bibitem{JAC15}
P. Jia, B. Andreotti and P. Claudin,
Paper waves in the wind.
Phys. Fluids \textbf{27}, 104101 (2015).

\bibitem{ZWBK17}
C. Zhang, J. Wang, W. Blake and J. Katz,
Deformation of a compliant wall in a turbulent channel flow.
J. Fluid Mech. \textbf{823}, 345-390 (2017).

\bibitem{SZ11}
M.J. Shelley and J. Zhang,
Flapping and bending bodies interacting with fluid flows.
Annu. Rev. Fluid Mech. \textbf{43}, 449-465 (2011).

\bibitem{S36}
A. Shields,
Anwendung der Aehnlichkeitsmechanik und der Turbulenzforschung auf die Geschiebebewegung
(Application of similarity principles and turbulence research to bedload motion),
Mitteilungen der Preu{\ss}ischen Versuchsanstalt f\"ur Wasserbau, Berlin \textbf{26}, 1-26  (1936).

\bibitem{CAC13}
F. Charru, B. Andreotti and P. Claudin,
Sand ripples and dunes.
Annu. Rev. Fluid Mech. \textbf{45}, 469-493 (2013).

\bibitem{LVMBMTOMDOA10}
M.Y. Louge, A. Valance, H. Mint Babah, J.-C. Moreau-Trouv\'e, A. Ould el-Moctar, P. Dupont, and D. Ould Ahmedou,
Seepage-induced penetration of water vapor and dust beneath ripples and dunes.
J. Geophys. Res. \textbf{115}, F02002 (2010).

\bibitem{MTLXB14}
R.A. Musa, S. Takarrouht, M.Y. Louge, J. Xu and M.E. Berberich,
Pore pressure in a wind-swept rippled bed below the suspension threshold.
J. Geophys. Res. \textbf{119}, 2475-2590 (2014).

\bibitem{S79}
R.B. Smith,
The influence of mountains on the atmosphere.
Adv. Geophys. \textbf{21}, 87-230  (1979).

\bibitem{W00}
C.D. Whiteman,
Mountain meteorology,
Oxford University Press (2000).

\bibitem{CWA13}
P. Claudin, G.F.S. Wiggs and B. Andreotti,
Field evidence for the upwind velocity shift at the crest of low dunes.
Boundary-Layer Meteorol. \textbf{148},195-206 (2013).

\bibitem{FRBA90}
J.J. Finnigan, M.R. Raupach, E.F. Bradley and G.K. Aldis,
A wind tunnel study of turbulent flow over a two-dimensional ridge.
Boundary-Layer Meteorol. \textbf{50}, 277-317 (1990).

\bibitem{GI89}
W. Gong and A. Ibbetson,
A wind tunnel study of turbulent flow over model hills.
Boundary-Layer Meteorol. \textbf{49}, 113-148 (1989).

\bibitem{GTD96}
W. Gong, P.A. Taylor and A. D\"ornbrack,
Turbulent boundary-layer flow over fixed aerodynamically rough two-dimensional sinusoidal waves.
J. Fluid Mech. \textbf{312}, 1-37 (1996).

\bibitem{K70}
J.M. Kendall,
The turbulent boundary layer over a wall with progressive surface waves.
J. Fluid Mech. \textbf{41}, 259-281 (1970).

\bibitem{M37}
H. Motzfeld,
Die turbulente Str\"omung an welligen W\"anden,
Zeit. Angewandte Math. Mech. \textbf{17}, 193-212 (1937).

\bibitem{WHCWWLC91}
W.S. Weng, J.C.R. Hunt, D.J. Carruthers, A. Warren, G.F.S. Wiggs, I. Linvingstone and I. Castro,
Air flow and sand transport over sand dunes.
Acta Mechanica \textbf{2}, 1-22 (1991).

\bibitem{LNDCRAFGC21}
P. L\"u, C. Narteau, Z. Dong, P. Claudin, S. Rodriguez, Z. An, L. Fernandez-Cascales, C. Gadal and S. Courrech du Pont,
Direct validation of dune instability theory.
Proc. Natl. Acad. Sci. USA \textbf{118}, e2024105118 (2021).

\bibitem{AH85}
J. Abrams and T.J. Hanratty,
Relaxation effects observed for turbulent flow over a wavy surface.
J. Fluid Mech. \textbf{151}, 443-455 (1985).

\bibitem{HSPZ17}
S.J. Haward, A.Q. Shen, J. Page, and T.A. Zaki,
Poisseuille flow over a wavy surface,
Phys. Rev. Fluids \textbf{12}, 124102 (2017).

\bibitem{NH01}
S. Nakagawa and T.J. Hanratty,
Particle image velocimetry measurements of flow over a wavy wall.
Phys. Fluids \textbf{13}, 3504 (2001).

\bibitem{NMcLW93}
J.M. Nelson, S.R. McLean and S.R. Wolfe,
Mean flow and turbulence fields over two-dimensional bed forms.
Water Resources Res. \textbf{29}, 3935-3953 (1993).

\bibitem{PKAR07}
D. Poggi, G.G. Katul, J.D. Albertson and L. Ridolfi,
An experimental investigation of turbulent flows over a hilly surface.
Phys. Fluids \textbf{19}, 036601 (2007).

\bibitem{V07}
J.G. Venditti,
Turbulent flow and drag over fixed two- and three-dimensional dunes.
J. Geophys. Res. \textbf{112}, F04008 (2007).

\bibitem{WN92}
P.L. Wiberg and J.M. Nelson,
Unidirectional flow over asymmetric and symmetric ripples.
J. Geophys. Res. \textbf{97}, 12745-12761 (1992).

\bibitem{ZCH77}
D.P. Zilker, G.W. Cook and T.J. Hanratty,
Influence of the amplitude of a solid wavy wall on a turbulent flow. Part 
1. Non-separated flows.
J. Fluid Mech. \textbf{82}, 29-51 (1977).

\bibitem{ZH79}
D.P. Zilker and T.J. Hanratty,
Influence of the amplitude of a solid wavy wall on a turbulent flow. Part 
2. Separated flows.
J. Fluid Mech. \textbf{90}, 257-271 (1979).

\bibitem{H81}
T.J. Hanratty,
Stability of surfaces that are dissolving or being formed by convective diffusion.
Annu. Rev. Fluid Mech. \textbf{13}, 231-252 (1981).

\bibitem{MJ10}
P. Meakin and B. Jamtveit,
Geological pattern formation by growth and dissolution in aqueous systems.
Proc. R. Soc. Lond. A \textbf{466}, 659-694 (2010).

\bibitem{CDA17}
P. Claudin, O. Dur\'an and B. Andreotti,
Dissolution instability and roughening transition.
J. Fluid Mech. \textbf{832}, R2 (2017).

\bibitem{BCBHMPCPD20}
M. Bordiec, S. Carpy, O. Bourgeois, C. Herny, M. Mass\'e, L. Perret, P. Claudin, S. Pochat and S. Dout\'e,
Sublimation waves: Geomorphic markers of interactions between icy planetary surfaces and winds.
Earth-Science Rev. \textbf{211}, 103350 (2020).

\bibitem{BRH01}
R. Bintanja, C.H. Reijmer and S.J. Hulscher,
Detailed observations of the rippled surface of Antarctic blue-ice areas.
J. Glaciol. \textbf{47}, 387-396 (2001).

\bibitem{T79}
R.M. Thomas,
Size of scallops and ripples formed by flowing water.
Nature \textbf{277}, 281-283 (1979).

\bibitem{DACW19}
O. Dur\'an, B. Andreotti, P. Claudin and C. Winter,
A unified model of ripples and dunes in water and planetary environments
Nature Geosci. \textbf{12}, 345-350 (2019). 

\bibitem{C70}
G.W. Cook,
Turbulent flow over solid wavy surfaces.
PhD thesis, University of Illinois at Urbana-Champaign (1970).

\bibitem{P00}
S.B. Pope,
Turbulent Flows.
Cambridge Univ. Press (2000).

\bibitem{vD56}
E.R. van Driest,
On turbulent flow near a wall.
J. Aeronaut. Sci. \textbf{23}, 1007-1011 (1956).

\bibitem{FS10}
K.A. Flack and M.P. Schultz,
Review of hydraulic roughness scales in the fully rough regime.
J. Fluid Eng. \textbf{132}, 041203 (2010).

\bibitem{SF09}
M.P. Schultz and K.A. Flack,
Turbulent boundary layers on a systematically varied rough wall.
Phys. Fluids \textbf{21}, 015104 (2009).

\bibitem{FH88}
K.A. Frederick and T.J. Hanratty,
Velocity measurements for a turbulent nonseparated flow over solid waves.
Exp. Fluids \textbf{6}, 477-86 (1988).

\bibitem{AFOKMC09}
B. Andreotti, A. Fourri\`ere, F. Ould-Kaddour, B. Murray and P. Claudin,
Giant aeolian dune size determined by the average depth of the atmospheric boundary layer.
Nature \textbf{457}, 1120-1123 (2009).

\bibitem{dALB97}
V. de Angelis, P. Lombardi and S. Banerjee,
Direct numerical simulation of turbulent flow over a wavy wall.
Phys. Fluids \textbf{9}, 2429-2442 (1997).

\bibitem{HS99}
D.S. Henn and R.I. Sykes,
Large-eddy simulation of flow over wavy surfaces.
J. Fluid Mech. \textbf{383}, 75-112 (1999).

\bibitem{MS94}
C. Maa\ss\ and U. Schumann,
Numerical simulation of turbulent flow over a wavy boundary. Direct and Large Eddy Simulation.
Editors P.R. Voke, L. Kleiser and J.P. Chollet, Kluwer, 287-297 (1994).

\bibitem{MS96}
C. Maa\ss\ and U. Schumann,
Direct numerical simulation of separated turbulent flow over a wavy boundary.
Flow Simulation with Higher Performance Computers. Editor E.H. Hirschel, 227-241 (1996).

\bibitem{SDB01}
M.V. Salvetti, R. Damiani and F. Beux,
Three-dimensional coarse large-eddy simulations of the flow above two-dimensional sinusoidal waves.
Int. J. Numer. Meth. Fluids \textbf{35}, 617-642 (2001).

\bibitem{MGCJH07}
F.J.R. Meysman, O.S. Galaktionov, P.L.M. Cook, F. Janssen and M. Huettel,
Quantifying biologically and physically induced flow and tracer dynamics in permeable sediments.
Biogeosciences \textbf{4}, 627-646 (2007).

\bibitem{LVOXHR13}
M.Y. Louge, A. Valance, A. Ould el-Moctar, J. Xu, A.G. Hay and R. Richer,
Temperature and humidity within a mobile barchan sand dune, implications for microbial survival.
J. Geophys. Res. \textbf{118}, 627-646 (2013).

\bibitem{CJVPA15}
P. Claudin, H. Jarry, G. Vignoles, M. Plapp and B. Andreotti,
Physical processes causing the formation of penitentes.
Phys. Rev. E \textbf{92}, 033015 (2015). 

\bibitem{TZGNBC04}
V. Tournat, V. Zaitsev, V. Gusev, V. Nazarov, P. B\'equin and B. Castagn\`ede,
Probing weak forces in granular media through nonlinear dynamic dilatancy: Clapping contacts and polarization anisotropy.
Phys. Rev. Lett. \textbf{92}, 085502 (2004). 

\bibitem{AFP13}
B. Andreotti, Y. Forterre and O. Pouliquen,
Granular Media, between Fluid and Solid.
Cambridge University Press (2013).

\bibitem{Vi07}
E. Villermaux,
Fragmentation.
Annu. Rev. Fluid Mech. \textbf{39}, 419-446 (2007). 

\end{thebibliography}
\end{document}